\documentclass[twocolumn,showkeys,aps,prb,showpacs]{revtex4-1}
\usepackage{graphicx}
\usepackage[CJKbookmarks,dvipdfm,colorlinks,linkcolor=blue,citecolor=blue]{hyperref}

\begin{document}

\title{Small strain induced large  piezoelectric coefficient in $\alpha$-AsP monolayer}

\author{San-Dong Guo,   Xiao-Shu Guo, Ya-Ying Zhang and Kui Luo}
\affiliation{School of Electronic Engineering, Xi'an University of Posts and Telecommunications, Xi'an 710121, China}
\begin{abstract}
Strain engineering can  effectively tune the electronic,  topological and piezoelectric properties of materials.
In this work, the small  strain (-4\% to 4\%) effects on  piezoelectric properties of $\alpha$-AsP monolayer are studied by
density functional theory (DFT).  The piezoelectric stress tensors $e_{ij}$ and elastic stiffness tensors $C_{ij}$  are reported by using density functional perturbation theory (DFPT) and finite difference method (FDM). It is found that the Young's modulus of $\alpha$-AsP monolayer shows
very strong anisotropy, and the armchair direction is very softer than zigzag direction, which provides possibility for tuning easily  piezoelectric coefficients along the armchair direction.
In considered strain range, uniaxial compressive  (tensile) strain along the armchair (zigzag) direction is found to raise observably both the $e_{22}$ and $d_{22}$ (absolute value). In fact, both compressive  strain along the armchair  direction and tensile strain along the zigzag direction
 essentially reduce the lattice constants along the armchair  direction, which can enhance the piezoelectric coefficients.
 The $e_{ij}$ of   $\beta$-AsP monolayer as a function of strain is also studied to illustrate the importance of particular puckered structure of $\alpha$-AsP in enhancing the  piezoelectric coefficients. A classic SnSe monolayer with  puckered structure is used to further declare that small strain along the armchair  direction can effectively improve the piezoelectric coefficients. For example, the $d_{22}$  of SnSe monolayer at -3.5\% strain is up to 628.8 pm/V  from unstrained 175.3 pm/V. For SnSe monolayer, a large peak is observed for $e_{22}$, which is due to a structural phase transition.  For $e_{16}$ of SnSe monolayer,  a large peak is also  observed  due to the cross of lattice constants  $a$ along the zigzag direction and $b$ along the armchair direction.
 A piezoelectric material  should  have a band gap for
prohibiting current leakage, and they all are semiconductors in considered strain range for all studied materials.
 Our works imply that small strain can effectually tune piezoelectric properties of materials with puckered structure, and can provide useful guidence for developing efficient nanopiezotronic devices.

.

\end{abstract}
\keywords{Piezoelectronics,  Monolayer,  Puckered structure}

\pacs{71.20.-b, 77.65.-j, 72.15.Jf, 78.67.-n ~~~~~~~~~~~~~~~~~~~~~~~~~~~~~~~~~~~Email:sandongyuwang@163.com}

\maketitle

\section{Introduction}
Since the discovery of graphene\cite{q0},  a series of two-dimensional
(2D)  materials  have been designed and fabricated.
Compared with  their bulk counterparts, these 2D materials exhibit unique
properties and application perspectives\cite{q1,q1-1,q2,q3}.
Many 2D monolayer materials are noncentrosymmetric, which can provide  opportunities for piezoelectric
applications, such as  sensors, actuators and
energy conversion devices\cite{q5,q6,q4}.
Experimentally, the piezoelectric coefficient ($e_{11}$=2.9$\times$$10^{-10}$ C/m)\cite{q5,q6} of monolayer  $\mathrm{MoS_2}$  has been reported, which is close to the previous theoretical
prediction\cite{q9}.
Recently, the Janus MoSSe monolayer has been experimentally  achieved by  replacing the top S atomic layer in  $\mathrm{MoS_2}$ with Se atoms, and the existence of vertical dipoles has been  proved, showing  an intrinsic vertical piezoelectric response\cite{q9-1}.
In theory, many 2D materials, including transition metal dichalchogenides (TMD), group IIA and IIB metal oxides, group III-V semiconductors, group-V binary semiconductors and Janus TMD, have been predicted to possess  piezoelectric properties \cite{q9,q7,q7-1,sn1,sn2,q11,q12,q13}.
Some of them possess high piezoelectric coefficient, like Janus $\mathrm{GaInS_2}$\cite{q13}, $\mathrm{In_2SSe}$\cite{q13} and CdO\cite{q9} with
coefficients $d_{11}$ of 8.33, 8.47 and 21.7 pm/V, which are
comparable and even higher than  other well-known bulk piezoelectric materials\cite{aln-1,aln-2,aln-3}.
 Moreover, the giant piezoelectricities of monolayer group IV monochalcogenides (SnSe,
SnS, GeSe and GeS) have been predicted,  which are as high as  75-251 pm/V \cite{sn1}for $d_{11}$ along the  the armchair  direction.
Beside an in-plane piezoelectricity,
additional out-of-plane piezoelectricity has been predicted by the first principle calculations in many  2D  materials, like Janus TMD\cite{q7,r2,r3}.
The simultaneous occurrence of semiconducting
and piezoelectric properties in these 2D materials may give rise to multifunctionality, which provides potential applications in piezotronics and piezophototronics\cite{w1,w2,w3}. However, the piezoelectric effect in most 2D materials is rather small, and thus new strategy should be proposed to
enhance  piezoelectric properties of some  2D materials with special structure.

\begin{figure}
  \includegraphics[width=8.0cm]{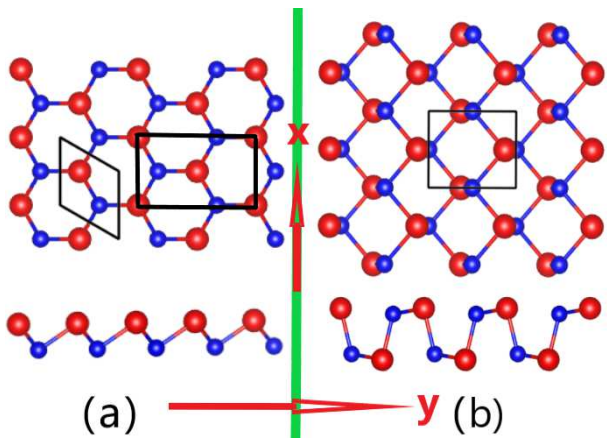}
  \caption{(Color online) The top view  and side view of crystal structure of   AsP monolayer for $\beta$- (a) and $\alpha$- (b) phases with armchair and zigzag being defined as y and x directions. The large red balls represent As atoms, and the  small blue balls for P atoms. The  rhombus primitive cell  and the rectangle supercell are shown for $\beta$-phase. }\label{t0}
\end{figure}

Although strain engineering has been widely  used to   effectively tune the electronic properties of 2D materials, strain-tuned their  piezoelectric properties are rarely reported\cite{r1}. Here, we use $\alpha$-AsP monolayer as an example to study strain effects on piezoelectric coefficients.
The $\alpha$-AsP monolayer is proposed as  2D solar cell donor with  1.54 eV direct band-gap
and mobility exceeding 14000 $\mathrm{cm^2V^{-1}s^{-1}}$\cite{q14}, and $\alpha$-phase few-layer of AsP has been
experimentally synthesized\cite{q15}. According to the structures (b) in \autoref{t0}, $\alpha$-phase
is soft along the armchair (y) direction with small elastic stiffness ($C_{22}$), which mean that  piezoelectric coefficients may be easily tuned.
Calculated results show that reduced lattice constants of $\alpha$-AsP monolayer along  armchair direction can  signally boost  the  $e_{22}$ and $d_{22}$. The strain effects on  $e_{ij}$ of   $\beta$-AsP monolayer are also investigated, and calculated results show the importance of particular puckered structure of $\alpha$-AsP in enhancing the  piezoelectric coefficients.
The SnSe monolayer as a  classic 2D material with puckered structure is used to illustrate that small strain along the armchair  direction can effectively improve the piezoelectric coefficients. For example,  a large peak can be observed for $e_{22}$ of SnSe monolayer at -3.5\% strain,
and the corresponding $d_{22}$ is 3.6 times of unstrained one. The underlying mechanism of a peak  for $e_{22}$  is due to a structural phase transition. A large peak of $e_{16}$ at -2\% strain is also observed,  which is due to the cross of $a$ and $b$.
Calculated results show that  they all are semiconductors in considered strain range for all studied materials, including monolayer  $\alpha$-AsP,  $\beta$-AsP  and SnSe.

The rest of the paper is organized as follows. In the next
section, we shall give our computational details and methods  about piezoelectric coefficients. In the third section, we shall present piezoelectric properties of monolayer  $\alpha$-AsP,  $\beta$-AsP  and SnSe as a function of strain. Finally, we shall give our discussion and conclusions in the fourth section.

\begin{figure*}
  \includegraphics[width=12.0cm]{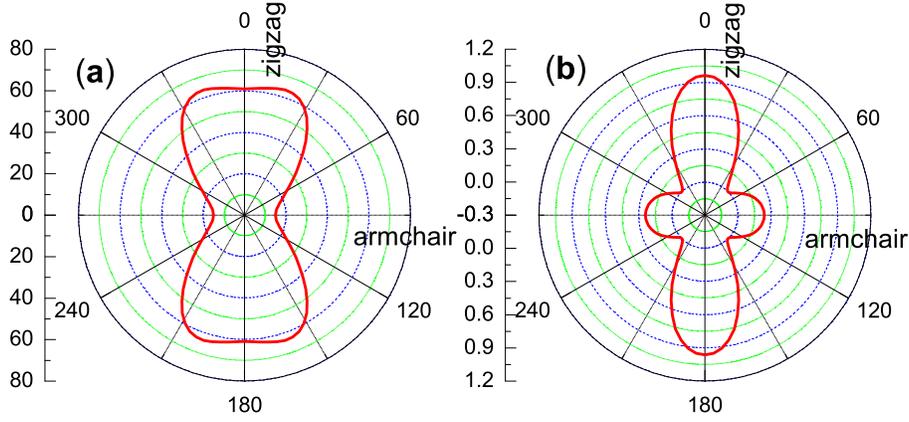}
  \caption{(Color online) The Young's modulus and Possion's ratio of $\alpha$-AsP monolayer as a function of the angle $\theta$.}\label{t1}
\end{figure*}

\section{Computational detail}
All the calculations are performed within DFT\cite{1} using the projected augmented wave
(PAW) method,  as implemented in the
plane wave, pseudopotential based VASP
Package\cite{pv1,pv2,pv3}. The exchange-correlation  functional
at the level of popular generalized gradient
approximation  of Perdew, Burke and  Ernzerhof  (GGA-PBE)\cite{pbe}
is employed in all our calculations.
 For studied monolayers, a vacuum spacing of
 more than 20 $\mathrm{{\AA}}$ along the z direction is included to avoid interactions
between two neighboring images. A kinetic cutoff energy of 500 eV is adopted for all investigated monolayers, and the total energy  convergence criterion is set
to $10^{-8}$ eV. The geometry optimization was considered
to be converged with the residual force on each atom being less than 0.0001 $\mathrm{eV.{\AA}^{-1}}$.
To obtain the piezoelectric strain coefficients $d_{ij}$, the elastic stiffness tensor $C_{ij}$ are calculated by using FDM, and the piezoelectric stress coefficients $e_{ij}$ are calculated by  DFPT method\cite{pv6}.
The 2D elastic coefficients $C^{2D}_{ij}$
 and   piezoelectric stress coefficients $e^{2D}_{ij}$
have been renormalized by the the length of unit cell along z direction ($Lz$):  $C^{2D}_{ij}$=$Lz$$C^{3D}_{ij}$ and $e^{2D}_{ij}$=$Lz$$e^{3D}_{ij}$.

\begin{figure*}
  \includegraphics[width=12cm]{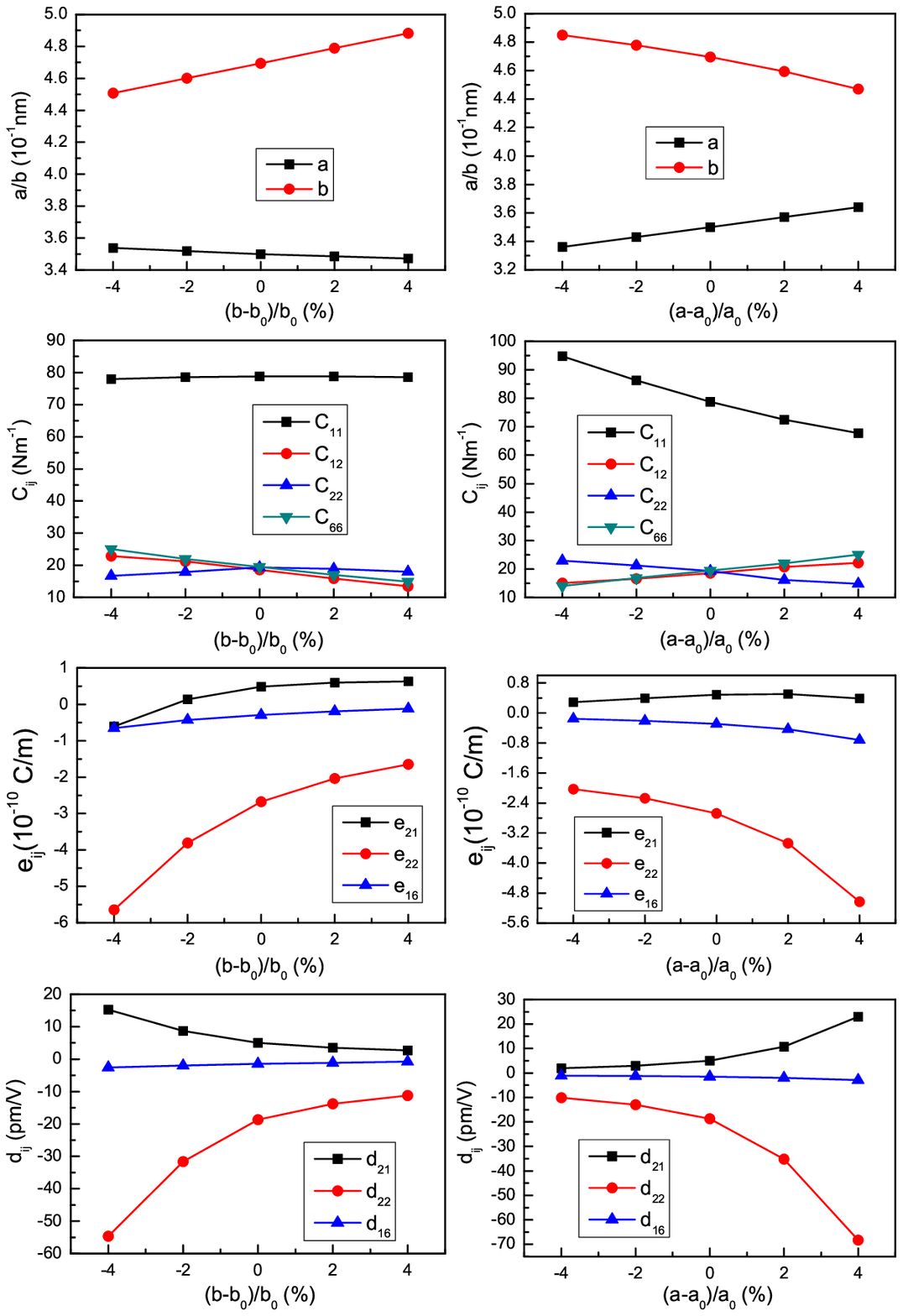}
\caption{(Color online) For monolayer $\alpha$-AsP, the lattice constants $a/b$, elastic constants $C_{ij}$, piezoelectric coefficients $e_{ij}$  and  $d_{ij}$  with the application of  uniaxial strain along the armchair
direction ($(b-b_0)/b_0$) and zigzag direction ($(a-a_0)/a_0$). }\label{t1-1}
\end{figure*}

\section{Piezoelectric properties}
Noncentrosymmetric crystals show a change of polarization under mechanical strain or stress, which can be described by the third-rank piezoelectric stress tensors  $e_{ijk}$ and strain tensor $d_{ijk}$. They  from the sum of ionic
and electronic contributions  can be expressed as:
 \begin{equation}\label{pe0}
      e_{ijk}=\frac{\partial P_i}{\partial \varepsilon_{jk}}=e_{ijk}^{elc}+e_{ijk}^{ion}
 \end{equation}
and
 \begin{equation}\label{pe0-1}
   d_{ijk}=\frac{\partial P_i}{\partial \sigma_{jk}}=d_{ijk}^{elc}+d_{ijk}^{ion}
 \end{equation}
In which $P_i$, $\varepsilon_{jk}$ and $\sigma_{jk}$ are polarization vector, strain and stress, respectively.
By employing Voigt notation and using the mapping of indices (11$\rightarrow$1,
22$\rightarrow$2, 33$\rightarrow$3, 23$\rightarrow$4, 31$\rightarrow$5 and 12$\rightarrow$6), the values of
$d_{ij}$ can be derived  using the relation:
 \begin{equation}\label{pe0-1}
   e_{ik}=d_{ij}C_{jk}
 \end{equation}
For 2D materials, we consider only in-plane strain components ($\varepsilon_{jk}$=$\sigma_{ij}$=0 for i=3 or j=3)\cite{q9,sn1,q11,q12}.
 The relation among elastic,  piezoelectric stress   and strain tensors  becomes:
 \begin{equation}\label{pe}
  \left(
    \begin{array}{ccc}
      e_{11} & e_{12} & e_{16} \\
     e_{21} & e_{22} & e_{26} \\
      e_{31} & e_{32} & e_{36} \\
    \end{array}
  \right)
  =
  \left(
    \begin{array}{ccc}
      d_{11} & d_{12} & d_{16} \\
      d_{21} & d_{22} & d_{26} \\
      d_{31} & d_{32} & d_{36} \\
    \end{array}
  \right)
    \left(
    \begin{array}{ccc}
      C_{11} & C_{12} & C_{16} \\
     C_{21} & C_{22} &C_{26} \\
      C_{61} & C_{62} & C_{66} \\
    \end{array}
  \right)
   \end{equation}
The piezoelectric stress tensors  $e_{ij}$  can be attained by DFPT,   and the elastic tensor $C_{ij}$ can be calculated by  FDM.
The piezoelectric  strain tensor $d_{ij}$ can be calculated by $e$ matrix multiplying  $C$ matrix inversion

Due to  the crystal symmetry, the number of independent
components can be reduced  in the elastic tensor,  piezoelectric stress and strain tensors.
The point group of $\alpha$-phase ($Pmn2_1$) is $mm2$. For 2D materials, there are three  nonzero piezoelectric constants: $e/d_{21}$, $e/d_{22}$ and $e/d_{16}$, and  there are five   nonzero elastic constants:
$C_{11}$, $C_{12}$=$C_{21}$, $C_{22}$ and $C_{66}$.  The   $d_{21}$, $d_{22}$ and $d_{16}$ are derived as:
\begin{equation}\label{pe2-7}
    d_{21}=\frac{e_{21}C_{22}-e_{22}C_{12}}{C_{11}C_{22}-C_{12}^2}
\end{equation}
\begin{equation}\label{pe2-7}
    d_{22}=\frac{e_{22}C_{11}-e_{21}C_{12}}{C_{11}C_{22}-C_{12}^2}
\end{equation}
\begin{equation}\label{e16}
    d_{16}=\frac{e_{16}}{C_{66}}
\end{equation}

Firstly, the lattice constants of $\alpha$-AsP are optimized, and the corresponding $a$=3.50 $\mathrm{\AA}$  and $b$=4.70 $\mathrm{\AA}$, which agree well with previous calculated values\cite{q14,q7-1}. To determine the piezoelectric strain tensors $d_{ij}$,  we first calculate
the elastic stiffness coefficients $C_{ij}$ and piezoelectric stress tensors $e_{ij}$, and then $d_{ij}$  can be attained.
These data are listed \autoref{tab-a}, along with previous theoretical values\cite{q7-1}.  The calculated $C_{ij}$ are very close to previous ones\cite{q7-1}.
On the basis of the elastic constants, the
Young's modulus $C_{2D}(\theta)$ and Poisson's ratio $\nu (\theta)$ along the in-plane  $\theta$ can be expressed as follows\cite{ela1}:
\begin{equation}\label{c2d}
C_{2D}(\theta)=\frac{C_{11}C_{22}-C_{12}^2}{C_{11}sin^4\theta+Asin^2\theta cos^2\theta+C_{22}cos^4\theta}
\end{equation}
\begin{equation}\label{c2d}
\nu (\theta)=\frac{C_{12}sin^4\theta-Bsin^2\theta cos^2\theta+C_{12}cos^4\theta}{C_{11}sin^4\theta+Asin^2\theta cos^2\theta+C_{22}cos^4\theta}
\end{equation}
where $A=(C_{11}C_{22}-C_{12}^2)/C_{66}-2C_{12}$ and $B=C_{11}+C_{22}-(C_{11}C_{22}-C_{12}^2)/C_{66}$
The calculated $C_{2D}(\theta)$  and $\nu (\theta)$ are plotted in \autoref{t1}.
It is found that  both  the Young's modulus  $C_{2D}(\theta)$ and Poisson's ratio
$\nu (\theta)$ show very strong mechanical anisotropy. A high Young's modulus
means that the material is rigid. Calculated results show that $\alpha$-AsP monolayer is very softer along the armchair than zigzag  direction, which means that strain can easily tune it's physical properties along armchair direction.
It is found that  the sign is different for $e/d_{21}$ and $e/d_{22}$ between our and previous ones\cite{q7-1}, which is due to the opposite y axes.
For $e/d_{22}$,  they are in good agreement, but they are not  very consistent for $e_{21}$, which may be due to different method.
In ref.\cite{q7-1} , the $e_{ij}$ coefficients are attained by evaluating the polarization
changes of a  unit cell under applied uniaxial strains, which is different from DFPT method. Here, the $e/d_{16}$ are also calculated, which don't be mentioned in ref.\cite{q7-1}.

\begin{table*}
\centering \caption{For monolayer  $\alpha$-AsP and SnSe,  the elastic constants $C_{ij}$ ($\mathrm{Nm^{-1}}$), piezoelectric coefficients $e_{ij}$ ($10^{-10}$C/m) and  $d_{ij}$ (pm/V).  The  previous calculated values are shown in  parentheses for $\alpha$-AsP\cite{q7-1} and SnSe\cite{sn1,sn2}.  }\label{tab-a}
  \begin{tabular*}{0.96\textwidth}{@{\extracolsep{\fill}}cccccc}
  \hline\hline
Name & $C_{11}$ & $C_{12}$& $C_{22}$&$C_{66}$& $e_{21}$ \\\hline\hline
AsP&78.76 (78.6)& 18.58 (18.4)       &19.31 (18.8)  & 19.47 & 0.485 (-0.25) \\
SnSe&42.82 (44.49\cite{sn1})& 18.89  (18.57\cite{sn1})  &23.06 (19.88\cite{sn1})  & 17.98&5.231 (10.8\cite{sn1}, 4.42\cite{sn2})\\\hline\hline
&&&&&\\\hline\hline
Name    & $d_{21}$& $e_{22}$&$d_{22}$& $e_{16}$&$d_{16}$\\\hline\hline
AsP    &  5.020 (-4.74)&-2.673 (2.68)& -18.670 (18.90)&  -0.288&  -1.481\\
SnSe&-65.108 (-80.31\cite{sn1})&28.135 (34.9\cite{sn1}, 24.18\cite{sn2})&175.315 (250.58\cite{sn1})&24.905 (28.17\cite{sn2})&138.502 \\\hline\hline
\end{tabular*}
\end{table*}
\begin{figure}
  \includegraphics[width=8cm]{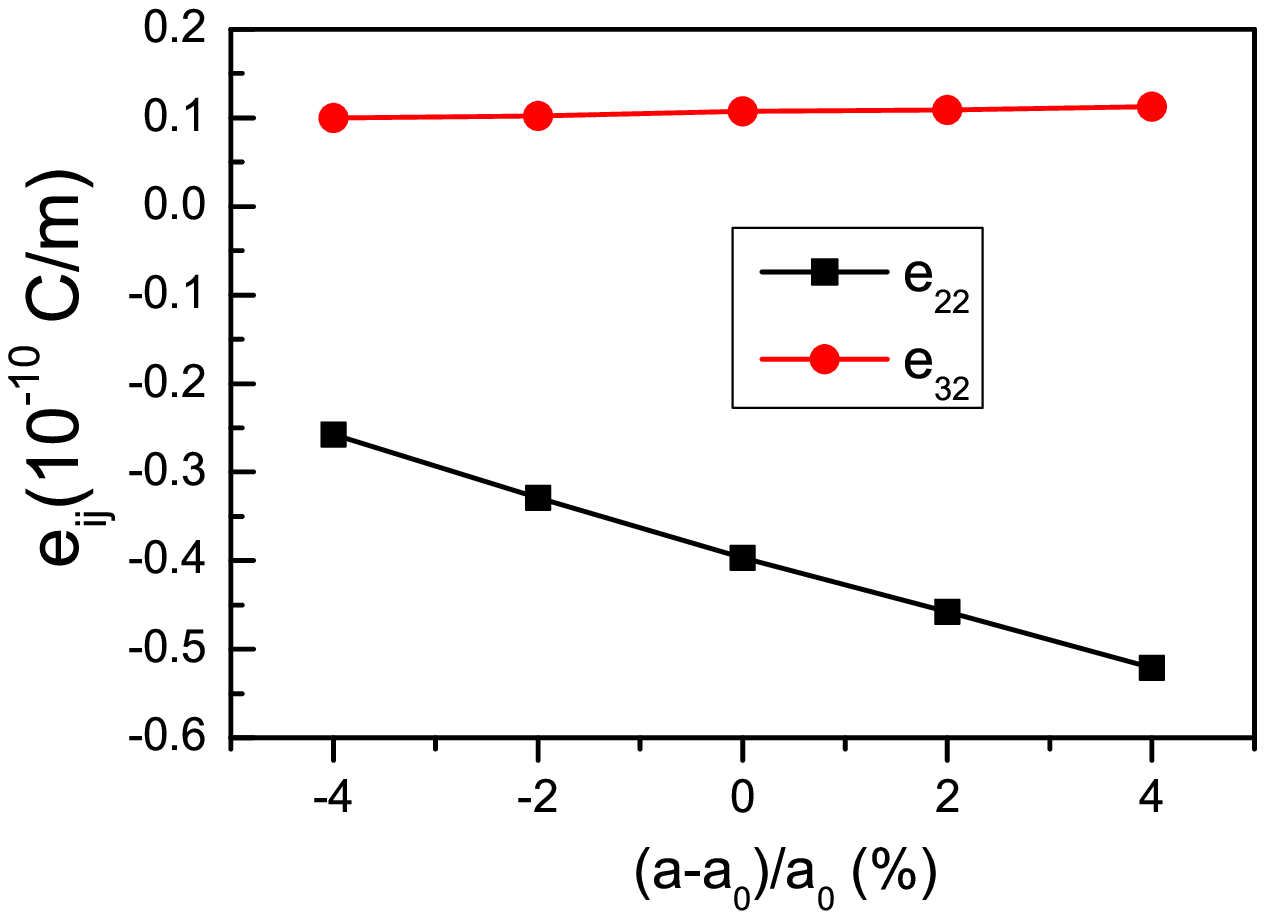}
\caption{(Color online) For monolayer $\beta$-AsP, the  piezoelectric coefficients $e_{ij}$   with the application of  biaxial strain. }\label{t1-2}
\end{figure}

Furthermore, to explore the strain effects on  piezoelectric coefficients of $\alpha$-AsP monolayer, the  uniaxial  strains along both  the armchair
direction  and zigzag direction are applied,  and the other unstrained direction is optimized fully.
 The lattice constants $a/b$, elastic constants $C_{ij}$, piezoelectric coefficients $e_{ij}$  and  $d_{ij}$  with the application of  uniaxial strain along the armchair
direction  and zigzag direction  are plotted \autoref{t1-1}. In considered strain range, $\alpha$-AsP monolayer always  shows very strong mechanical anisotropy between the armchair
direction  and zigzag direction due to very larger $C_{11}$ than $C_{22}$. When the uniaxial  strain is applied  along  the armchair
direction,  the $C_{11}$ has little dependence on strain. However, the $C_{11}$ shows a  monotone decrease with  the application of  uniaxial strain along  zigzag direction.  The $C_{22}$,  $C_{12}$ and  $C_{16}$ show opposite  monotonic trend with the application of  uniaxial strain between the armchair direction  and zigzag direction, when the strain changes from -4\% to 4\%.

It is found that  piezoelectric
coefficient $d_{22}$ (absolute value) and $d_{21}$ increase with strain from 4\% to -4\% along  armchair
direction. However, it is opposite with the application of  uniaxial strain along  zigzag direction, and the $d_{22}$ (absolute value) and $d_{21}$ increase from -4\%  to 4\% strain.  Especially for $d_{22}$,  a large enhancement
can be observed. At
4\% uniaxial compressive  (tensile) strain along the armchair (zigzag) direction, the $d_{22}$ (absolute value) reaches up to 54.663 pm/V (68.315 pm/V) from unstrained 18.670 pm/V. In fact, when both  the armchair
direction  or zigzag direction  is subjected to  uniaxial strain from 4\% to -4\%,  the lattice constants $b$ along the armchair  direction  decreases essentially. It is clearly seen that the enhancement for $d_{22}$  mainly depends on $e_{22}$, and the  similar strain dependence between $e_{22}$ and $d_{22}$ can be seen from \autoref{t1-1}.  At
-4\% (4\%) strain along the armchair (zigzag) direction, the $e_{22}$ (absolute value) reaches up to 5.642 $10^{-10}$C/m (5.029 $10^{-10}$C/m) from unstrained 2.673 $10^{-10}$C/m.
For $d_{16}$, the  small dependability on strain is observed for both armchair
direction  and zigzag direction, which is because the $e_{16}$ (absolute value) and $C_{66}$ have the similar monotonous dependability on strain,
 giving rise to little change for $d_{16}$ according to \autoref{e16}.

  \begin{figure}
  \includegraphics[width=8cm]{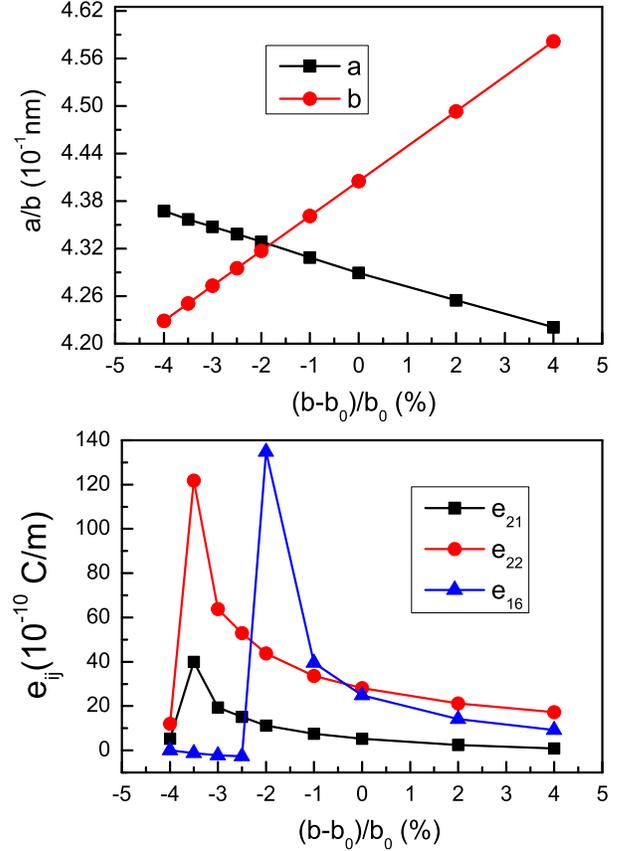}
\caption{(Color online)For monolayer SnSe, the lattice constants $a/b$ and piezoelectric coefficients $e_{ij}$  with the application of  uniaxial strain along the armchair
direction ($(b-b_0)/b_0$). }\label{t1-3}
\end{figure}

The particular puckered structure of $\alpha$-AsP should   be very important for enhancement of $e_{22}$, when the strain is applied.
To confirm that, the piezoelectric stress coefficients of monolayer $\beta$-AsP  are studied as a function of strain.
The geometric structure of the monolayer $\beta$-AsP is plotted (a) in \autoref{t0}, which  has a graphene-like buckled honeycomb structure.
Firstly, we perform symmetry analysis for $\beta$-AsP.
Due to a $3m$ point-group symmetry of $\beta$-phase ($P3m_1$),
  five nonzero piezoelectric constants ($e/d_{22}$=-$e/d_{21}$=-$e/d_{16}$ and $e/d_{32}$=$e/d_{31}$) can be attained, and  there are five   nonzero elastic constants: $C_{11}$=$C_{22}$, $C_{12}$=$C_{21}$ and  $C_{66}$=($C_{11}$-$C_{12}$)/2.  The   $d_{22}$ and $d_{32}$ from $e_{ij}$ can be expressed as:
\begin{equation}\label{pe2-7}
    d_{22}=\frac{e_{22}}{C_{22}-C_{21}}
\end{equation}
\begin{equation}\label{pe2-7}
    d_{32}=\frac{e_{32}}{C_{22}+C_{21}}
\end{equation}
The optimized lattice constants $a$=$b$=3.44 $\mathrm{\AA}$ for $\beta$-AsP, which agrees well with previous values\cite{q14,q7-1}.
The calculated elastic stiffness coefficients $C_{ij}$ and piezoelectric  tensors $e/d_{ij}$  are shown in \autoref{tab-b}, along with available calculated values\cite{q7-1}. It is found that our calculated  $C_{21}$  is  twice as large as one from ref.\cite{q7-1}.  To further identify this, the elastic stiffness coefficients $C_{ij}$ and piezoelectric  tensors $e/d_{ij}$ of $\beta$-SbAs are also calculated, which are listed in \autoref{tab-b}.
The similar result is observed for  $C_{21}$ of $\beta$-SbAs.
 As with $\alpha$-AsP, some similar comparisons between our results and ones from ref.\cite{q7-1}  can be attained for  $\beta$-AsP. Experimentally, it is convenient for $\beta$-AsP to apply biaxial strain due to hexagonal symmetry.
The $e_{ij}$ as a function of  biaxial strain (-4\%-4\%) are plotted in \autoref{t1-2}. In considered strain range, it is clearly seen that the $e_{22}$ (absolute value) increases linearly from -4\% to 4\%, and the $e_{32}$ has  small dependability on strain.
 At 4\% strain, the $e_{22}$ (absolute value) reaches up to 0.521 $10^{-10}$C/m  from unstrained 0.397 $10^{-10}$C/m,  increased by 0.3 times,
  which is smaller than one  of  $\alpha$-AsP (0.9 times at -4\% strain along armchair  direction  and  1.1 times at 4\% strain along zigzag direction). Thus, the piezoelectric coefficients of a  material with the particular puckered structure  should sensitivity depends on strain.

\begin{figure}
  \includegraphics[width=8.0cm]{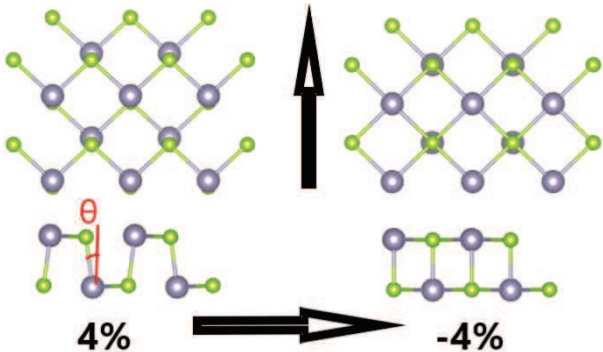}
  \caption{(Color online) The top view (001)  and side view (010) of crystal structure of  $\alpha$-SnSe at 4\% and -4\% strain,  The arrows show armchair  direction. }\label{t2}
\end{figure}
\begin{figure}
  \includegraphics[width=8.0cm]{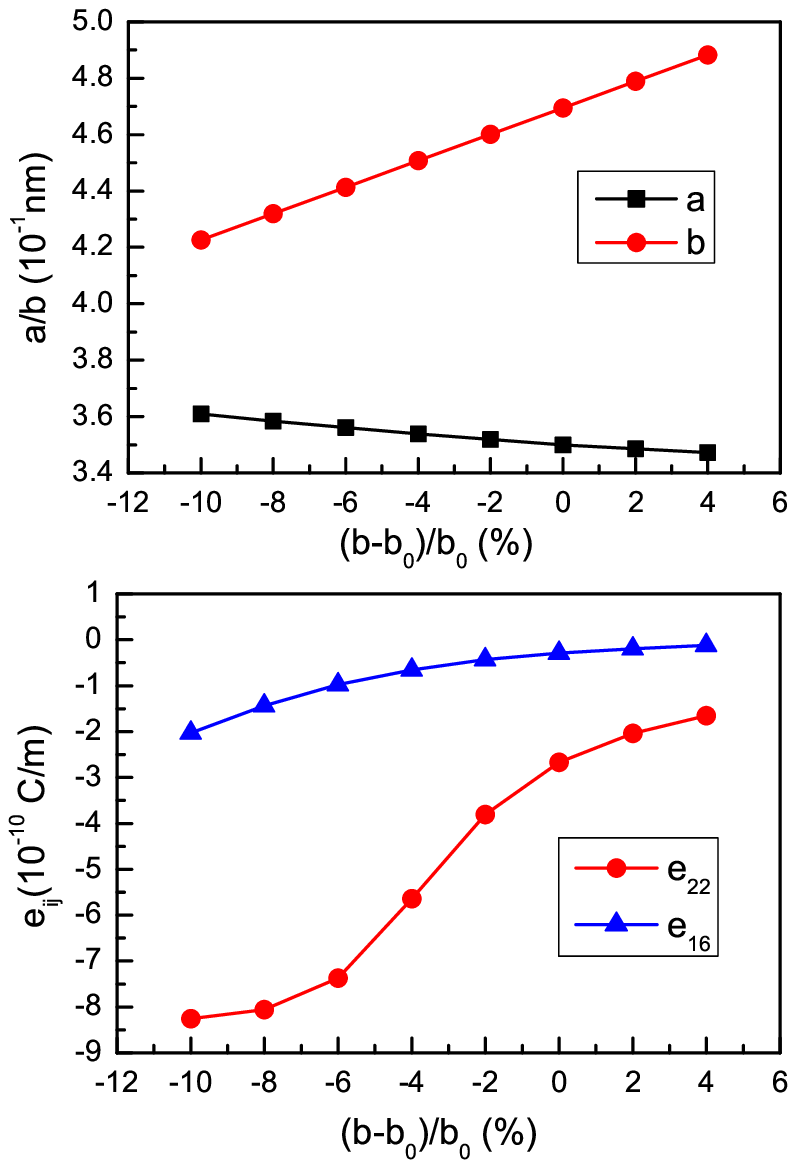}
  \caption{(Color online) For monolayer $\alpha$-AsP, the lattice constants $a/b$ and piezoelectric coefficients $e_{ij}$  with the application of  uniaxial strain along the armchair
direction ($(b-b_0)/b_0$)[-10\% to 4\%]. }\label{t3}
\end{figure}

\begin{table*}
\centering \caption{For monolayer  $\beta$-AsP and $\beta$-SbAs,  the elastic constants $C_{ij}$ ($\mathrm{Nm^{-1}}$), piezoelectric coefficients $e_{ij}$ ($10^{-10}$C/m) and  $d_{ij}$ (pm/V).  The  previous calculated values\cite{q7-1} are shown in  parentheses.}\label{tab-b}
  \begin{tabular*}{0.96\textwidth}{@{\extracolsep{\fill}}ccccccc}
  \hline\hline
Name & $C_{22}$ & $C_{21}$& $e_{22}$ & $d_{22}$& $e_{32}$&$d_{32}$\\\hline\hline
AsP&65.21 (62.9)& 19.52 (9.5) &-0.397 (0.36)& -0.868 (0.67) &0.108 (0.007)& 0.127 (0.01)\\
SbAs&47.54 (40.8)&17.42 (8.0) &-0.575 (0.54) &-1.909 (1.65) &-0.018 (-0.014) & -0.028   (-0.029)\\\hline\hline
\end{tabular*}
\end{table*}
\begin{figure}
  \includegraphics[width=8cm]{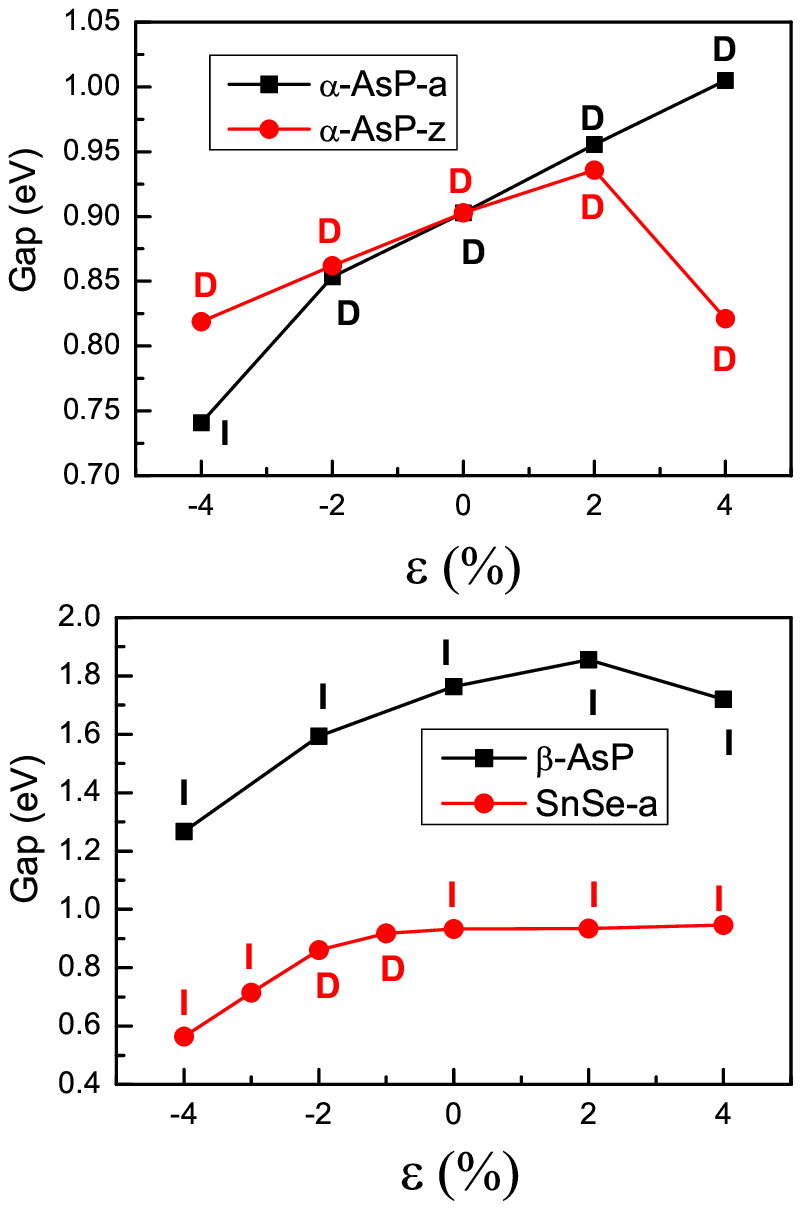}
\caption{(Color online)The gap as a function of strain for $\alpha$-AsP monolayer along the armchair  direction ($\alpha$-AsP-a) and  the zigzag direction ($\alpha$-AsP-z),   $\beta$-AsP monolayer ($\beta$-AsP) and SnSe monolayer along the armchair  direction (SnSe-a). The 'I' and 'D' mean indirect and direct gap, respectively.}\label{t1-g}
\end{figure}

Small strain along armchair
direction can effectively improve the piezoelectric effect of a 2D material with  puckered structure. To prove that,  the piezoelectric coefficients of monolayer SnSe with  the application of uniaxial strain along the armchair
direction are investigated. The monolayer SnSe has the same crystal structure with $\alpha$-AsP, and  few layer
SnSe has been fabricated in  experiment\cite{sn3}.  Our optimized lattice constants $a$=4.29 $\mathrm{\AA}$  and $b$=4.41 $\mathrm{\AA}$, which are close to previous calculated values ($a$=4.24 $\mathrm{\AA}$, $b$=4.35 $\mathrm{\AA}$\cite{sn1} and $a$=4.30 $\mathrm{\AA}$, $b$=4.36 $\mathrm{\AA}$\cite{sn2}). The calculated elastic stiffness coefficients $C_{ij}$ and piezoelectric  tensors $e/d_{ij}$ of SnSe monolayer  are listed in \autoref{tab-a}, along with available theoretical values\cite{sn1,sn2}. It is found that our calculated $C_{ij}$ are very close to previous values\cite{sn1}, and the $e_{21}$ and $e_{22}$  are between ones in ref.\cite{sn1} and ones  in ref.\cite{sn2}. The difference may be due to    different calculated method and lattice constants.
The SnSe monolayer is
subjected to  uniaxial  strain  along the armchair
direction, and the lattice constants $a/b$ and piezoelectric coefficients $e_{ij}$  as a function of strain are plotted in \autoref{t1-3}.
In considered strain range, it is clearly seen that  all $e_{ij}$ have a rise, and then rapidly fall, when the strain changes from 4\% to -4\%.
A huge enhancement  is observed  for all $e_{ij}$ at small strain.  The $e_{22}$ is up to 121.847  $10^{-10}$C/m at -3.5\% strain, and the $e_{16}$
reaches up to 134.831  $10^{-10}$C/m at -2\% strain, and the $e_{21}$ for 39.885 $10^{-10}$C/m at  -3.5\% strain. With respect to unstrained ones, the $e_{22}$, $e_{16}$ and $e_{21}$ are   increased by 3.33, 4.41  and 6.63 times, respectively. Thus,  the $e_{ij}$  has very sensitive dependence on strain in a material with puckered structure [(b) in \autoref{t0}]. The key  $d_{22}$  of SnSe monolayer at -3.5\% strain  can be improved to 628.814 pm/V  from unstrained 175.315 pm/V.

A large peak for $e_{22}$  of SnSe monolayer is  observed, and the underlying mechanism is structural phase transition.
Around -3.5\% strain,  the symmetry changes  from  $Pmn2_1$  to $Pmmn$ (ignoring calculation error), and the crystal structures of  $\alpha$-SnSe at 4\% and -4\% strain are shown \autoref{t2}. It is clearly seen that the angle  $\theta$ nearly becomes  zero with strain from 4\% to -4\%, which means a structural phase transition. In fact, when the compressive strain increases continuously, the  symmetry of SnSe monolayer strictly becomes $Pmmn$, and the piezoelectric effect disappears due to centrosymmetry. For  $e_{16}$, a large peak is also observed at -2\% strain, and then the $e_{16}$  is almost zero
 with increasing compressive strain. It is noted that the peak and  cross point of $a$ and $b$ coincide at -2\% strain. It is concluded that the $e_{16}$  disappears, when lattice constants ($b$) along armchair direction is less than lattice constants ($a$) along zigzag direction. To prove above these, the $\alpha$-AsP monolayer is
subjected to  uniaxial  strain  along the armchair
direction, increased to -10\% for compressive strain, and  the lattice constants $a/b$ and piezoelectric coefficients $e_{ij}$  with the application of  uniaxial strain along the armchair
direction (-10\% to 4\%) are plotted \autoref{t3}. It is clearly seen that $a$ and $b$ of $\alpha$-AsP do not have cross point, and the $e_{16}$ (absolute value) monotonically increase.  The $e_{22}$ (absolute value) has also a monotonous increase, and it is found that no structural phase transition is produced.

\section{Discussions and Conclusion}
A 2D material,  exhibiting  piezoelectricity,  should   break inversion symmetry,  and then has a band gap.
To confirm studied materials to be semiconductors at applied strain,  the GGA gaps as a function of strain for $\alpha$-AsP monolayer along the armchair  direction and  the zigzag direction,   $\beta$-AsP monolayer  and SnSe monolayer along the armchair  direction  are plotted in \autoref{t1-g}.
Our calculated  unstrained gaps  for monolayer  $\alpha$-AsP (0.90 eV),  $\beta$-AsP (1.76 eV)  and SnSe (0.93 eV) agree well with previous
theoretical values\cite{q14,qs}. It is clearly seen that, in considered strain range, all studied 2D materials have a band gap.
For $\alpha$-AsP monolayer along the armchair  direction, the gap monotonically increases from -4\% to 4\%, and it changes from indirect gap to direct one. However, for $\alpha$-AsP monolayer along the zigzag  direction, the gap firstly increases, and then decreases, but the direct band gap does not change. For  $\beta$-AsP monolayer with biaxial strain, the gap shows an up-and-down trend, and they in considered strain range all are indirect gap.
For monolayer SnSe along the armchair  direction, the gap increases rapidly, and basically remain unchanged. The  gap  property changes from indirect gap to direct one to indirect one.

In summary, the strain effects on  piezoelectric properties of $\alpha$-AsP monolayer are studied by using first-principles calculations. Calculated results show that the uniaxial compressive (tensile) strain applied along the armchair  (zigzag) direction are very effective in improving $e/d_{22}$.
Essentially, reducing the lattice constants along the armchair  direction can enhance the piezoelectric coefficients.
By studying $e_{ij}$ of   $\beta$-AsP monolayer as a function of strain,  the particular puckered structure of $\alpha$-AsP  is very important to enhance the  piezoelectric coefficients.  A classic puckered SnSe monolayer is used to prove that  small strain along the armchair  direction can effectively improve the piezoelectric coefficients. For example, the $d_{22}$  of SnSe monolayer at -3.5\% strain can be improved to 628.8 pm/V  from unstrained 175.3 pm/V.  A structural phase transition  can induce a large peak of $e_{22}$ for SnSe monolayer, and a cross of $a$ and $b$ can lead to a large peak of $e_{16}$.
Our works imply that the piezoelectric properties of a 2D material with puckered structure can be easily tuned by strain (The similar phenomenon may be observed for monolayer GeS, GeSe and SnS with puckered structure.), and can provide
new idea for designing  nanopiezotronic devices  by controlling
the conversion of mechanical to electrical energy.

\begin{acknowledgments}
This work is supported by the Natural Science Foundation of Shaanxi Provincial Department of Education (19JK0809). We are grateful to the Advanced Analysis and Computation Center of China University of Mining and Technology (CUMT) for the award of CPU hours and WIEN2k/VASP software to accomplish this work.
\end{acknowledgments}


\begin{references}
\bibitem{q0}K. S.  Novoselov et al.,  Science \textbf{306}, 666	(2004).

\bibitem{q1}C. Lee, X. D.  Wei, J. W. Kysar and  J. Hone, Science  \textbf{321}, 385  (2008).

\bibitem{q1-1}M.  Xu, T.  Liang, M.  Shi and H. Chen,  Chem. Rev. \textbf{113}, 3766 (2013).

\bibitem{q2}H. L. Zhuang, R. G. Hennig,  Phys. Rev. B \textbf{88}, 115314 (2013).


\bibitem{q3}H. Park, A.  Wadehra, J. W.  Wilkins and  A. H. Castro Neto,  Appl. Phys. Lett.
\textbf{100}, 253115 (2012).

\bibitem{q5} W. Wu, L. Wang, Y. Li, F. Zhang, L. Lin, S. Niu, D. Chenet,
X. Zhang, Y. Hao, T. F. Heinz, J. Hone and Z. L. Wang,
Nature \textbf{514}, 470 (2014).


\bibitem{q6}H. Zhu, Y. Wang, J. Xiao, M. Liu, S. Xiong, Z. J. Wong, Z. Ye,
Y. Ye, X. Yin and X. Zhang, Nat. Nanotechnol. \textbf{10},
151 (2015).
\bibitem{q4}W. Wu and Z. L. Wang, Nat. Rev. Mater. \textbf{1}, 16031 (2016).

\bibitem{q9}M. N. Blonsky, H. L. Zhuang, A. K. Singh and R.  G. Hennig,  ACS Nano, \textbf{9},
9885 (2015).

\bibitem{q9-1}A. Y. Lu, H. Zhu, J. Xiao, C. P. Chuu, Y. Han, M. H. Chiu,
C. C. Cheng, C. W. Yang, K. H. Wei, Y. Yang, Y. Wang,
D. Sokaras, D. Nordlund, P. Yang, D. A. Muller, M. Y. Chou,
X. Zhang and L. J. Li, Nat. Nanotechnol. \textbf{12}, 744 (2017).

\bibitem{q7}L. Dong, J. Lou and V. B. Shenoy, ACS Nano, \textbf{11},
8242 (2017).


\bibitem{q7-1}H. B. Yin, J. W. Gao, G. P. Zheng, Y. X. Wang and Y. C. Ma, J. Phys. Chem. C \textbf{121}, 25576 (2017).


\bibitem{sn1}R. X. Fei, We. B. Li, J. Li and L. Yang, Appl. Phys. Lett.  \textbf{107}, 173104 (2015)



\bibitem{sn2}L. C. Gomes, A. Carvalho  and A. H. Castro Neto, Phys. Rev. B  \textbf{92}, 214103 (2015).

\bibitem{q11}K. N. Duerloo, M. T. Ong and E. J. Reed, J. Phys. Chem. Lett. \textbf{3}, 2871 (2012).


\bibitem{q12}Y. Chen,  J. Y. Liu,  J. B. Yu,  Y. G. Guo and Q. Sun, Phys. Chem. Chem. Phys.
 \textbf{21}, 1207 (2019).



\bibitem{q13}Y. Guo, S. Zhou, Y. Bai, J. Zhao,  Appl. Phys. Lett.\textbf{110}, 163102 (2017).
\bibitem{aln-1} C. M. Lueng, H. L. Chang, C. Suya and C. L. Choy, J. Appl. Phys. \textbf{88},
 5360 (2000).

\bibitem{aln-2} A. Hangleiter, F. Htzel, S. Lahmann and U. Rossow,  Appl. Phys. Lett.
\textbf{83},  1169 (2003).

\bibitem{aln-3} S. Muensit, E. M. Goldys and I. L. Guy,  Appl. Phys. Lett. \textbf{75},
3965 (1999).


\bibitem{r2}Dimple, N. Jena, A. Rawat, R. Ahammed,
M. K. Mohanta and A. D. Sarkar,  J. Mater. Chem. A, \textbf{6},
24885 (2018).

\bibitem{r3}S. D. Guo, X. S. Guo, R. Y. Han and Y. Deng,  Phys. Chem. Chem. Phys. (2019). DOI: 10.1039/C9CP04590B.

\bibitem{w1} W. Wu, L. Wang, Y. Li, F. Zhang, L. Lin, S. Niu, D. Chenet,
X. Zhang, Y. Hao, T. F. Heinz, J. Hone and Z. L. Wang,
Nature \textbf{514}, 470 (2014).


\bibitem{w2} W. Wu and Z. L. Wang, Nat. Rev. Mater. \textbf{1}, 16031 (2016).

\bibitem{w3}Z. L. Wang, Adv. Mater. \textbf{24}, 4632 (2012).

\bibitem{r1}N. Jena, Dimple, S. D.  Behere  and A. D. Sarkar, J. Phys. Chem. C  \textbf{121}, 9181 (2017).

\bibitem{q14}M. Q. Xie, S. L. Zhang, B.  Cai et al., Nano Energy  \textbf{28},  433 (2016).


\bibitem{q15}B. Liu, M. Kopf, A. N. Abbas, X. Wang, Q. Guo, Y. Jia, F. Xia, R. Weihrich,
F. Bachhuber, F. Pielnhofer, et al., Adv. Mater. \textbf{27},  4423 (2015).

\bibitem{1}P. Hohenberg and W. Kohn, Phys. Rev. \textbf{136},
B864 (1964); W. Kohn and L. J. Sham, Phys. Rev. \textbf{140},
A1133 (1965).

\bibitem{pv1} G. Kresse, J. Non-Cryst. Solids \textbf{193}, 222 (1995).

\bibitem{pv2} G. Kresse and J. Furthm$\ddot{u}$ller, Comput. Mater. Sci. 6, \textbf{15} (1996).

\bibitem{pv3} G. Kresse and D. Joubert, Phys. Rev. B \textbf{59}, 1758 (1999).

\bibitem{pbe}J. P. Perdew, K. Burke and M. Ernzerhof, Phys. Rev. Lett. \textbf{77}, 3865 (1996).


 \bibitem{pv6}X. Wu, D. Vanderbilt and  D. R.  Hamann, Phys. Rev. B  \textbf{72}, 035105 (2005).








\bibitem{ela1}E. Cadelano, P. L. Palla, S. Giordano and L. Colombo,  Phys. Rev. B  \textbf{82}, 235414 (2010).


\bibitem{sn3}L. Li, Z. Chen, Y. Hu, X. Wang, T. Zhang, W. Chen, and Q. Wang, J. Am.
Chem. Soc.  \textbf{135}, 1213 (2013).


\bibitem{qs}L. C. Gomes and A. Carvalho, Phys. Rev. B \textbf{92}, 085406 (2015).

\end{references}
\end{document}